\documentclass[twocolumn,aps,pra,showpacs,nofootinbib]{revtex4-1}

\usepackage{graphicx}
\usepackage{amsmath,amssymb}
\usepackage{longtable}

\def\endproof{\vrule height6pt width6pt depth0pt}

\begin{document}


\title{All-versus-nothing proofs with $n$ qubits distributed between $m$ parties}


\author{Ad\'{a}n Cabello}
 \email[]{adan@us.es}
 \affiliation{Departamento de F\'{\i}sica Aplicada II, Universidad
 de Sevilla, E-41012 Sevilla, Spain}

\author{Pilar Moreno}
 \email[]{mpmoreno@us.es}
 \affiliation{Departamento de F\'{\i}sica Aplicada II, Universidad
 de Sevilla, E-41012 Sevilla, Spain}


\date{\today}




\begin{abstract}
All-versus-nothing (AVN) proofs show the conflict between
Einstein, Podolsky, and Rosen's elements of reality and the
perfect correlations of some quantum states. Given an $n$-qubit
state distributed between $m$ parties, we provide a method with
which to decide whether this distribution allows an $m$-partite
AVN proof specific for this state using only single-qubit
measurements. We apply this method to some recently obtained
$n$-qubit $m$-particle states. In addition, we provide all
inequivalent AVN proofs with less than nine qubits and a
minimum number of parties.
\end{abstract}


\maketitle


\section{Introduction}
\label{Sec:1}


Einstein, Podolsky, and Rosen (EPR) showed that quantum
mechanics is incomplete, in the sense that not every element of
reality has a counterpart inside the theory \cite{EPR35}. EPR
proposed the following criterion to identify an element of
reality: ``If, without in any way disturbing a system, we can
predict with certainty (i.e., with probability equal to unity)
the value of a physical quantity, then there exists an element
of physical reality corresponding to this physical quantity''
\cite{EPR35}. In practice, nondisturbance can be guaranteed
when the measurements are performed on distant systems.
Predictions with certainty are possible for states having
perfect correlations. A quantum state $\rho$ has $p$ perfect
correlations when there are $p$ different observables $O_i$
such that $\langle O_i \rangle_{\rho}=1$.

Thirty years after EPR's paper, Bell proved that there is a
irresoluble conflict between EPR's elements of reality and
quantum mechanics \cite{Bell64}. All-versus-nothing (AVN)
proofs are the most direct way to reveal this conflict. An AVN
proof is based on a set of $s$ perfect correlations of a
specific quantum state. The name ``all-versus-nothing''
\cite{Mermin90b} reflects one particular feature of these
proofs: If one assumes EPR elements of reality, then $s-q$ of
these perfect correlations lead to a conclusion that is the
opposite of the one obtained from a subset of the other $q$
perfect correlations. If all $s$ correlations are essential to
obtain a contradiction (i.e., if the contradiction vanishes
when we remove one of them), then the AVN proof is said to be
critical.

The first AVN proof was obtained by Heywood and Redhead
\cite{HR83}. However, the most famous AVN proof is Greenberger,
Horne, and Zeilinger's (GHZ) \cite{GHZ89, GHSZ90, Mermin90a}.
The first bipartite AVN proof with qubits is in Refs.
\cite{Cabello01a, Cabello01b}. The first bipartite AVN proof
with qubits and using only single-qubit measurements is in
Refs. \cite{Cabello05a, Cabello05b}. The interest of the case
in which the parties are restricted to perform single-qubit
measurements is motivated by the practical difficulty of making
general $N$-qubit measurements ($N \ge 2$) when the qubits are
encoded in different degrees of freedom of the same particle.

Recently, several $n$-qubit $m$-particle states ($n > m$)
having perfect correlations have been experimentally prepared,
for instance, 4-qubit two-photon \cite{VPMDB07}, 6-qubit
two-photon \cite{BLPK05, CVDMC09}, 6-qubit four-photon
\cite{GYXGCLPYCP09}, 8-qubit four-photon \cite{GLYXGGCPCP08},
and 10-qubit five-photon graph states \cite{GLYXGGCPCP08}.

For these $n$-qubit $m$-particle states, a natural problem is
the following: Consider $m$ distant parties; party $i$ can
perform single-qubit measurements on particle $i$, particle $i$
contains $n_i \ge 1$ qubits ($\sum_{i=1}^m n_i = n$); which
$n$-qubit $m$-particle states allow $m$-partite AVN proofs?
This problem has been solved for the case of $m=2$
particles/parties \cite{CM07}. In this article we address the
problem for an arbitrary number $m$ of particles/parties.

The article is organized as follows: In Sec.~\ref{Sec:2}, we
show that there is an equivalence between pure states allowing
AVN proofs and graph states. This will simplify the task of
finding all inequivalent $n$-qubit $m$-partite AVN proofs.

An $m$-partite AVN proof is specific for an $n$-qubit
$m$-particle graph state when there is no graph state with
fewer qubits satisfying the same correlations. In
Sec.~\ref{Sec:3} we discuss the requirements of an $m$-partite
AVN proof to be specific for an $n$-qubit $m$-particle graph
state, and describe a method to decide whether a given
$n$-qubit $m$-particle graph state allows a specific
$m$-partite AVN proof. We apply this method to decide whether
some $n$-qubit $m$-particle graph states recently prepared in
the laboratory allow $m$-partite AVN proofs. As supplementary
material \cite{SM}, we provide a computer program to decide
whether a given $n$-qubit $m$-particle graph state allows a
specific $m$-partite AVN proof.

In Sec.~\ref{Sec:4} we solve the following problem: Given an
$n$-qubit graph state, what is the minimum number $m$ of
parties that allows a specific $m$-partite AVN proof. As
supplementary material (See Footnote 1), we provide a computer
program to obtain, given an $n$-qubit graph state, all
distributions between $m$ parties and all distributions between
a minimum number of parties which allow AVN proofs.

The solution of the previous problem allows us to obtain all
inequivalent distributions allowing AVN proofs since any
distribution obtained from one allowing a specific AVN proof by
giving qubits that originally belong to the same party to new
parties will also allow an AVN proof. As supplementary
material, we provide all inequivalent distributions between a
minimum number $m$ of parties allowing specific $m$-partite AVN
proofs for all $n$-qubit graph states of $n \le 8$
qubits.\footnote{See EPAPS Document No. $\cdots$ for all
inequivalent distributions between a minimum number $m$ of
parties allowing specific $m$-partite AVN proofs for all
$n$-qubit graph state of $n \le 8$ qubits.}


\section{AVN proofs and graph states}
\label{Sec:2}


\subsection{AVN proofs and stabilizer states}


An AVN proof requires an $n$-qubit quantum state distributed
between $m$ parties. This state has a set of perfect
correlations between the results of single-qubit measurements.
These correlations must satisfy two requirements. First, they
must allow us to define $m$-partite EPR's elements of reality.
This means that every single-qubit observable involved in the
AVN proof must satisfy EPR's criterion of elements of reality
(i.e., its value can be predicted with certainty using only the
results of single-qubit measurements on distant particles).
Second, they must lead to a contradiction when EPR's criterion
of elements of reality is assumed. Therefore the conclusion of
an AVN proof is that if the quantum predictions are correct,
observables which satisfy EPR's condition cannot have
predefined results since it is impossible to assign them values
which simultaneously satisfy the perfect correlations predicted
by quantum mechanics.

Perfect correlations are necessary to establish elements of
reality and to prove that they are incompatible with quantum
mechanics. Therefore the states which allow AVN proofs must be
simultaneous eigenstates of a sufficient number of commuting
$n$-fold tensor products of single-qubit operators. Indeed, the
following observations lead us to the conclusion that without
loss of generality, we can restrict our attention to a
particular family of states.

Two different single-qubit operators $A$ and $B$ on the same
qubit cannot commute. A necessary condition to make $n$-fold
tensor products be commuting operators is to choose $A$ and $B$
to be anticommuting operators. Therefore, in an AVN proof, all
single-qubit operators corresponding to the same qubit must be
anticommuting operators. The maximum number of anticommuting
single-qubit operators is three. Therefore, without loss of
generality, we can restrict our attention to a specific set of
three single-qubit anticommuting operators on each qubit, for
example the Pauli matrices $X=\sigma_x$, $Y=\sigma_y$, and
$Z=\sigma_z$. This leads us to the concept of stabilizer state.
An $n$-qubit stabilizer state is the simultaneous eigenstate
with eigenvalue $1$ of a set of $n$ independent commuting
elements of the Pauli group (i.e., the group, under matrix
multiplication, of all $n$-fold tensor products of $X$, $Y$,
$Z$ and the identity $\openone$). The $n$ independent elements
are called stabilizer generators and generate a maximally
Abelian subgroup, the stabilizer group of the state
\cite{Gottesman96}. The $2^n$ elements of the stabilizer group
are the stabilizing operators and provide all the perfect
correlations of the stabilizer state.

A further simplification is possible since any stabilizer state
is local Clifford equivalent (i.e., equivalent under the local
unitary operations that map the Pauli group to itself under
conjugation) to a graph state \cite{VDM04}. Therefore the
problem of which $n$-qubit pure states and distributions of
qubits between the parties allow $m$-partite AVN proofs is
reduced to the problem of which $n$-qubit graph states and
distributions allow $m$-partite AVN proofs.


\subsection{Graph states}


A graph state \cite{HEB04} is a stabilizer state whose
generators can be written with the help of a graph. $|G\rangle$
is the $n$-qubit state associated with the graph $G$, which
gives a recipe both for preparing $|G\rangle$ and for obtaining
$n$ stabilizer generators that uniquely determine $|G\rangle$.
On one hand, $G$ is a set of $n$ vertices (each representing a
qubit) connected by edges (each representing an Ising
interaction between the connected qubits). On the other hand,
the stabilizer generator $g_i$ is obtained by looking at the
vertex $i$ of $G$ and the set $N(i)$ of vertices which are
connected to $i$ and is defined by
\begin{equation}
g_i = X_i \otimes_{j \in N(i)} Z_j, \label{grule}
\end{equation}
where $X_i$, $Y_i$, and $Z_i$ denote the Pauli matrices acting on
the $i$th qubit. $|G\rangle$ is the only $n$-qubit state that
fulfills
\begin{equation}
g_i |G\rangle = |G\rangle \text{ for }i=1,\ldots,n. \label{erule}
\end{equation}
Therefore the stabilizer group is
\begin{equation}
S(|G\rangle)=\{s_j, j=1,\ldots,2^n\}, \;\;\; s_j=\prod_{i \in
I_j(G)} g_i, \label{stabilizer}
\end{equation}
where $I_j(G)$ denotes a subset of
$\left\{g_i\right\}_{i=1}^N$. The stabilizing operators provide
all the perfect correlations of $|G\rangle$:
\begin{equation}
\langle G| s_j |G\rangle = 1. \label{stabilizingoperatorseqs}
\end{equation}
Graph states associated with connected graphs have been
exhaustively classified. There is only 1 two-qubit graph state
(equivalent to a Bell state), only 1 three-qubit graph state
(equivalent to a GHZ state), and 2 four-qubit graph states
(equivalent to a GHZ and a cluster state), 4 five-qubit graph
states, 11 six-qubit graph states, 26 seven-qubit graph states
\cite{HEB04}, and 101 eight-qubit graph states \cite{CLMP09}.


\section{$n$-qubit $m$-partite AVN proofs}
\label{Sec:3}


\subsection{Specific $m$-partite AVN proofs}


The perfect correlations of any graph state associated with a
connected graph of three or more vertices lead to
contradictions with the concept of elements of reality when
each qubit is distributed to a different party \cite{DP97,
SASA05, HDERVB06, CM07}. However, the first problem consists of
finding whether these contradictions are specific to a given
distribution of a graph state or, on the contrary, they can be
obtained with a graph state of fewer qubits.

For example, take the four-party AVN proof based on the
following four perfect correlations of the distribution of the
four-qubit fully connected graph state $|\text{FC}_4\rangle$ (a
four-qubit GHZ state) in which each qubit belongs to a
different party:
\begin{subequations}
\begin{align}
X_1 Z_2 Z_3 Z_4 = 1, \\
Z_1 X_2 Z_3 Z_4 = 1, \\
Z_1 Z_2 X_3 Z_4 = 1, \\
-X_1 X_2 X_3 Z_4 = 1.
\end{align}
\end{subequations}
This is an example of an AVN proof which is nonspecific for the
state $|\text{FC}_4\rangle$, the reason being that neither the
contradiction nor the definition of the elements of reality
involved in this contradiction requires any choice from the
party which has the fourth qubit. This party only has to
measure $Z_4$ and broadcast the result. The only role of the
result of $Z_4$ is to guarantee that $X_1$, $Z_1$, $X_2$,
$Z_2$, $X_3$, and $Z_3$ are elements of reality in a four-party
scenario. However, the contradiction occurs for any result of
$Z_4$. It occurs because the following equations cannot be
simultaneously satisfied:
\begin{equation}
X_1 Z_2 Z_3 = Z_1 X_2 Z_3 = Z_1 Z_2 X_3 = -X_1 X_2 X_3.
\end{equation}
The particular value of $Z_4$ is irrelevant. The same
contradiction can be obtained using the perfect correlations of
a three-qubit fully connected graph state $|\text{FC}_3\rangle$
(a three-qubit GHZ state) distributed between three parties.

The next example illustrates that whether an AVN proof is
specific can depend on the way in which the qubits are
distributed between the parties. Consider the AVN proof based
on the following four correlations of the four-qubit linear
cluster state $|\text{LC}_4\rangle$ associated with the graph
where qubit $1$ is connected to qubit $2$, which is connected
to qubit $3$, which is connected to qubit $4$:
\begin{subequations}
\begin{align}
Y_1 Y_2 Z_3 = 1, \\
Z_1 X_2 Z_3 = 1, \\
Z_1 Y_2 Y_3 Z_4 = 1, \\
-Y_1 X_2 Y_3 Z_4 = 1.
\end{align}
\end{subequations}

If the qubits are distributed so that each qubit goes to a
different party, then the AVN proof is not specific since the
party who has the fourth qubit does not need to make any
choice, neither for the contradiction nor for the definition of
the elements of reality. The contradiction
\begin{equation}
Y_1 Y_2 Z_3 = Z_1 X_2 Z_3 = Z_1 Y_2 Y_3 = -Y_1 X_2 Y_3
\end{equation}
can be obtained from the perfect correlations of a three-qubit
linear cluster state $|\text{LC}_3\rangle$ associated with the
graph where qubit $1$ is connected to qubit $2$, which is
connected to qubit $3$.

However, if qubits $1$ and $4$ belong to Alice, and qubits $2$
and $3$ belong to Bob, then the only way to guarantee that, for
example, $X_2$ is an element of reality in this scenario (i.e.,
that its result can be predicted using only the results of
measurements on Alice's side) is by using the following perfect
correlation of the $|\text{LC}_4\rangle$:
\begin{equation}
Z_1 X_2 X_4=1.
\end{equation}
Therefore the party who has qubit $4$ must choose between at
least two measurements. To sum up, an AVN proof is specific for
a given distribution of a graph state when at least two
observables of all the qubits are involved.

Since the additional correlations needed to define the elements
of reality can (together with those already used for the
contradiction) involve additional contradictions, it is
appropriate that the observables needed to guarantee that other
observables are elements of reality (like $X_4$ and $Z_4$ in
the previous example) are themselves elements of reality.
Therefore hereinafter we will focus on AVN proofs in which at
least two of the observables of all the qubits are elements of
reality. It can be easily seen that when two Pauli observables,
for example, $X_i$ and $Y_i$, are elements of reality, then the
third Pauli observable, $Z_i$, is also an element of reality.
Therefore we shall focus only on those graph states and
distributions in which the three Pauli observables of each and
every one of the qubits are elements of reality.


\subsection{When does a distribution allow a specific AVN proof?}


The next problem is, given a distribution of an $n$-qubit graph
state between $m$ parties, how to decide whether it is one in
which all single-qubit Pauli observables are elements of
reality. For that purpose, it is useful to note that the $2^n$
perfect correlations (i.e., stabilizing operators) of an
$n$-qubit graph state can be classified in four classes:

1. There are $2^{n-2}$ stabilizing operators (i.e., a quarter
of the stabilizing operators of the graph state) that allow us
to predict $X_i$ from the results of measurements on other
qubits: those that are products of the stabilizer generator
$g_i$ [defined in Eq. (\ref{grule})], an even number
(hereinafter ``even'' includes zero) of $g_j$ with $j \in
N(i)$, and an arbitrary number (hereinafter ``arbitrary
number'' includes zero) of $g_k$ with $k \neq i$ and $k \not
\in N(i)$.

2. There are $2^{n-2}$ stabilizing operators that allow us to
predict $Y_i$ from the results of measurements on other qubits:
those that are products of $g_i$, an odd number of $g_j$ with
$j \in N(i)$, and an arbitrary number of $g_k$ with $k \neq i$
and $k \not \in N(i)$.

3. There are $2^{n-2}$ stabilizing operators that allow us to
predict $Z_i$ from the results of measurements on other qubits:
those that are products of an odd number of $g_j$ with $j \in
N(i)$ and an arbitrary number of $g_k$ with $k \neq i$ and $k
\not \in N(i)$.

4. There are $2^{n-2}$ stabilizing operators that contain
$\openone_i$: those that are products of an even number of
$g_j$ with $j \in N(i)$ and an arbitrary number of $g_k$ with
$k \neq i$ and $k \not \in N(i)$.

Each particle can carry more than one qubit. It is therefore
convenient to denote as $P(i)$ the set of qubits which are in
the same particle as qubit $i$. The previous classification of
the stabilizing operators is useful in the following sense:
Given the distribution of an $n$-qubit graph state between $m$
parties, $X_i$ is an element of reality if and only if there
exists a stabilizing operator of the graph state which
satisfies the following two requirements: (1) It does not
contain $g_j$ for all $j \in P(i)$ but contains an even number
of $g_k$ with $k \in N(j)$ and (2) it contains $g_i$ and an
even number of $g_l$ with $l \in N(i)$. For instance, consider
the four-qubit linear cluster state $|\text{LC}_4\rangle$
associated with the graph where qubit $1$ is connected to qubit
$2$, which is connected to qubit $3$, which is connected to
qubit $4$, distributed such that Alice has qubits 1 and 4 and
Bob has qubits 2 and 3. The question is, is $X_1$ an element of
reality? This is equivalent to the question, is there a
stabilizing operator such that it does not contain $g_4$ [since
$P(1)=\{4\}$] but contains an even number (necessarily zero) of
$g_3$ [since $N(4)=\{3\}$] and $g_1$ and an even number
(necessarily zero) of $g_2$ [since $N(1)=\{2\}$]? The answer is
yes; the only stabilizing operator with these properties is
$g_1=X_1 Z_2$.

Similarly, $Y_i$ is an element of reality if and only if there
is a stabilizing operator satisfying (1) and the following
condition: (3) It contains $g_i$ and an odd number of $g_l$
with $l \in N(i)$.

Finally, $Z_i$ is an element of reality if and only if there is
a stabilizing operator satisfying (1) and the following
condition: (4) It does not contain $g_i$ but contains an odd
number of $g_l$ with $l \in N(i)$.

To decide whether a specific distribution allows a specific AVN
proof, we first focus on qubit $i$ and test whether $X_i$ and
$Y_i$ are elements of reality. If either is not an element of
reality, then the distribution does not allow a specific AVN
proof. If both are elements of reality, then we test whether
$X_j$ and $Y_{j}$ of qubit $j$ are elements of reality, and so
on for all the qubits. If all $X_i$ and $Y_i$ are elements of
reality, then the distribution allows a specific AVN proof.

Indeed, there are simple cases where it can easily be seen that
a distribution does not allow an AVN proof. For example, if
more than $n/2$ qubits are carried by the same particle, for
qubits of the particle with more than $n/2$ qubits, either
requirement (1) is incompatible with (2), or (1) is
incompatible with (3) and (4). An alternative proof will be
provided in Sec.~\ref{Sec:4}. If there is a qubit $i$ such that
$N(i) \in P(i)$ (i.e., if in the graph representing the state,
qubit $i$ is connected only to qubits of the same particle),
requirement (1) is incompatible with requirements (3) and (4).
As supplementary material (see Footnote 1), we provide a
computer program to decide whether a given $n$-qubit
$m$-particle graph state allows a specific $m$-partite AVN
proof.


\subsection{Examples}


As an example of the application of these rules, it is
interesting to discuss whether some recently prepared 6-qubit
two- and four-particle states allow specific AVN proofs,
assuming the natural scenario in which each party has one
particle.


\begin{figure}[t]
\center
\includegraphics[width=0.86\linewidth]{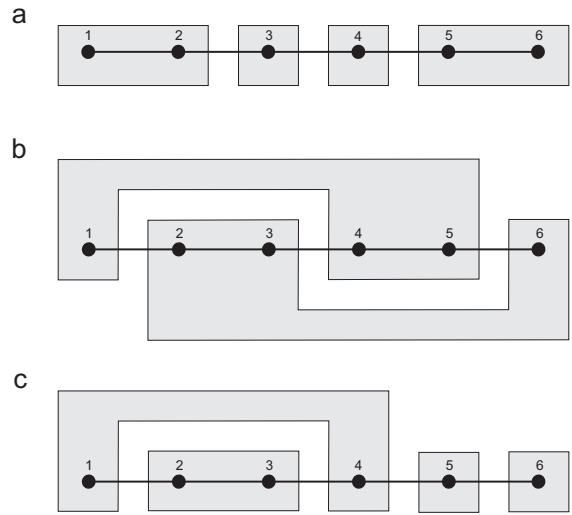}
\caption{Different distributions of a six-qubit linear cluster state
$|\text{LC}_6\rangle$ between two and four particles. Each gray area
represents a particle. (a) corresponds to the
four-photon state prepared in Ref. \cite{GYXGCLPYCP09}.
(b) corresponds to the two-photon state prepared in Ref.
\cite{CVDMC09}. In (a), not all single-qubit Pauli
observables are EPR elements of reality, and therefore no AVN proof
is possible. (b) and (c) allow AVN proofs.}
\label{Fig1}
\end{figure}


Figure~\ref{Fig1} contains several possible distributions of a
six-qubit linear cluster state $|\text{LC}_6\rangle$.
Figure~\ref{Fig1}(a) represents the four-photon
$|\text{LC}_6\rangle$ prepared in Ref. \cite{GYXGCLPYCP09}.
This distribution does not allow a specific AVN proof since
qubit 1 is connected only to qubit 2 and qubit 6 is connected
only to qubit 5.

Figure~\ref{Fig1}(b) represents the two-photon
$|\text{LC}_6\rangle$ prepared in Ref. \cite{CVDMC09}. This
distribution satisfies all the requirements thus allows a
specific AVN proof. Indeed, Fig.~\ref{Fig1} (b) represents the
only bipartite distribution of the six-qubit linear cluster
state which allows a specific AVN proof \cite{CM07}. Some
distributions of $|\text{LC}_6\rangle$ in four particles
allowing AVN proofs can be trivially obtained from
Fig.~\ref{Fig1}(b) by splitting qubits that belong to the same
particle into several particles. For instance, a distribution
allowing a specific AVN proof is illustrated in
Fig.~\ref{Fig1}(c). It can be easily seen that there is no
distribution in four particles which allows a specific AVN
proof which cannot be obtained from the distribution in
Fig.~\ref{Fig1}(b).


\begin{figure}[t]
\center
\includegraphics[width=0.86\linewidth]{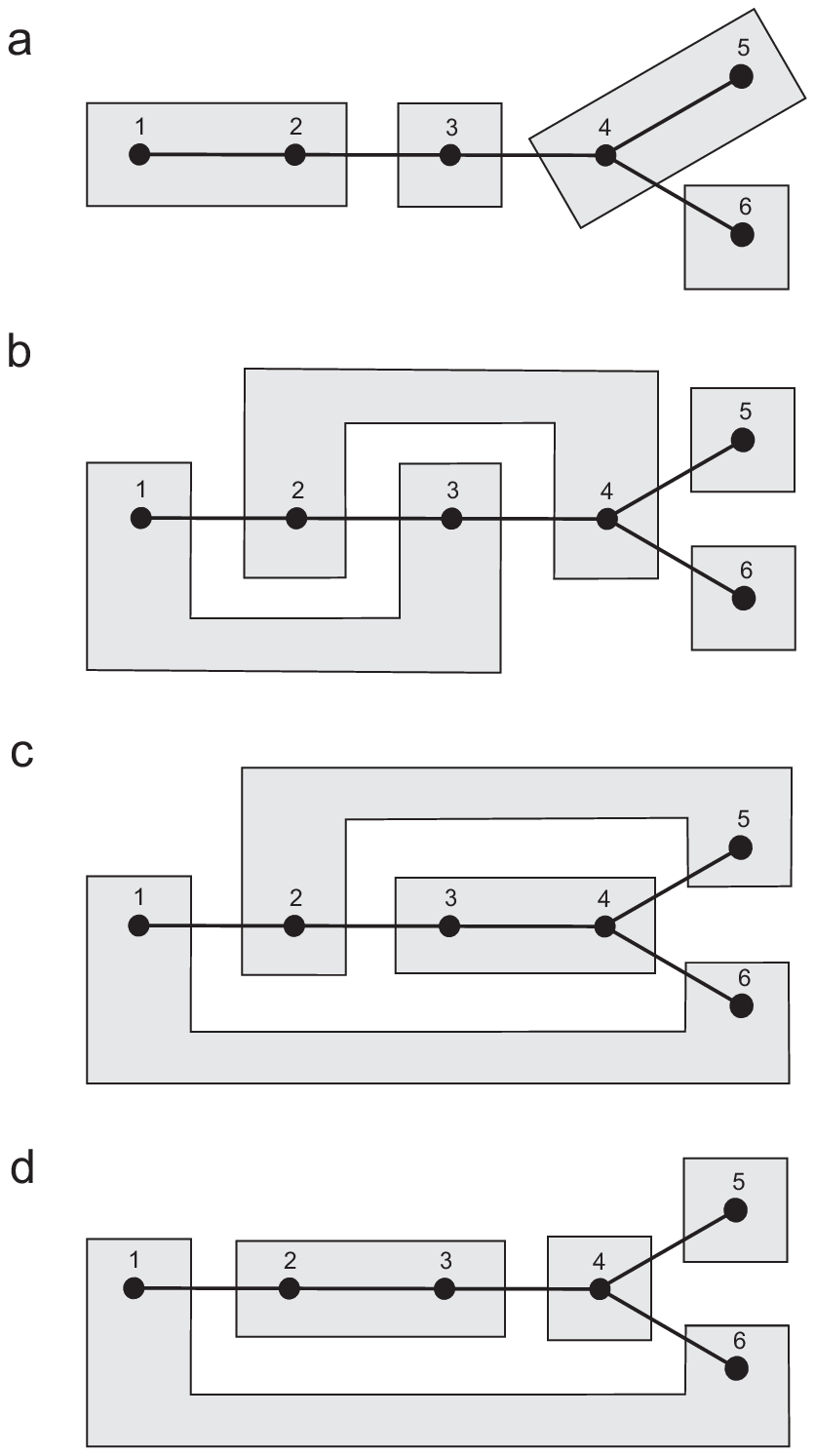}
\caption{Different distributions of the six-qubit Y graph state
between four particles. (a) corresponds to the state
prepared in Ref. \cite{GYXGCLPYCP09} and does not allow a specific
AVN proof. (b)--(d) allow specific AVN proofs.}
\label{Fig2}
\end{figure}


Figure~\ref{Fig2} contains several possible distributions of a
six-qubit $Y$-graph state $|Y_6\rangle$. Figure~\ref{Fig2}(a)
represents the four-photon $|Y_6\rangle$ prepared in Ref.
\cite{GYXGCLPYCP09}. This distribution does not allow a
specific AVN proof since qubit 1 is connected only to qubit 2
and qubit 5 is connected only to qubit 4.
Figures.~\ref{Fig2}(b)--(d) represent distributions of
$|Y_6\rangle$ between four particles allowing specific AVN
proofs.


\section{AVN proofs with a minimum number $m$ of parties} \label{Sec:4}


\subsection{Possible distributions between a minimum number of
parties}


In the previous section, we have seen that $|Y_6\rangle$ admits
specific AVN proofs when their qubits are suitably distributed
between four particles. The question is whether $|Y_6\rangle$
admits specific AVN proofs when it is distributed between three
particles or less, or more generally speaking, the question is,
given an $n$-qubit graph state, what is the minimum number of
parties $m$ which allows $m$-partite AVN proofs specific for
this state?

The following definition will be useful for solving this problem.
Let us define the reduced stabilizer of particle $A$'s qubits as the
one obtained from the stabilizer of the original state by replacing
the observables on all other particles' qubits with identity
matrices.


{\em Lemma:} A distribution of $n =n_\text{max}+n_B+\cdots+n_m$
qubits between $m$ parties such that $n_\text{max} \ge n_B \ge
\ldots \ge n_m$ allows $m$-partite elements of reality if and
only if $n_\text{max} \le n_B + \cdots + n_m$.


{\em Proof:} Suppose that particle $m_i$ carries qubits
$1,\ldots, n_\text{max}$, where $n_\text{max}$ is the maximum
number of qubits carried by any particle, and that particle
$m_j$ carries qubits $n_\text{max}+1,\ldots, n_\text{max}+n_j$.
If $X_1$, $Y_1$, $Z_1$, $X_2,\ldots, Z_{n_\text{max}}$ are
elements of reality, then the reduced stabilizer of $m_i$'s
qubits must contain
\begin{subequations}
\begin{align}
&X_1\otimes \openone_2\otimes\cdots\otimes \openone_{n_\text{max}},\label{eq1}\\
&Y_1\otimes \openone_2\otimes\cdots\otimes \openone_{n_\text{max}},\label{eq2}\\
&Z_1\otimes \openone_2\otimes\cdots\otimes \openone_{n_\text{max}},\label{eq3}\\
&\openone_1\otimes X_2\otimes\ldots\otimes \openone_{n_\text{max}},\ldots,\label{eq6}\\
&\openone_1\otimes \openone_2\otimes\ldots\otimes Z_{n_\text{max}}.\label{eq7}
\end{align}
\end{subequations}
Moreover, the reduced stabilizer of $m_i$'s qubits must contain
all possible products of Eqs. (\ref{eq1})--(\ref{eq7}), that
is, all possible variations with repetition of the four
elements $\openone$, $X$, $Y$, and $Z$, choosing $n_i$, which
are $4^{n_\text{max}}=2^{2n_\text{max}}$. A similar reasoning
applies to the three Pauli matrices of each and every one of
$m_j$'s qubits. Therefore the reduced stabilizer of $m_j$'s
qubits must also contain all possible products of
\begin{subequations}
\begin{align}
&X_{n_\text{max}+1}\otimes \openone_{n_\text{max}+2}\otimes\ldots\otimes \openone_{n_\text{max}+n_j},\ldots, \\
&\openone_{n_\text{max}+1}\otimes \openone_{n_\text{max}+2}\otimes\ldots\otimes Z_{n_i+n_j},
\end{align}
\end{subequations}
which are $4^{n_j}=2^{2n_j}$. However, the reduced stabilizer
of the sum of the parties $m_i$ and $m_j$ has only
$2^{n_\text{max} + n_j}$ terms; therefore the only possibility
is that $n_\text{max} = n_j$.\hfill\endproof


Given an $n$-qubit graph state, $n_\text{max}$ restricts the
possible minimum numbers of particles and the possible numbers
of qubits per particle. Given $n$, Table~\ref{Table1} presents
the possible minimum numbers of particles and the corresponding
possible distributions. Other possible distributions are
already contained between those in Table~\ref{Table1}, but in
those cases the number of particles is not the minimum.


\begin{table}[htb]
\caption{\label{Table1}Possible distributions of an $n$-qubit
graph state between a minimum number $m$ of particles. For
instance, (2,2,1) denotes a distribution of $n=5$ qubits
between $m=3$ particles such that particles 1 and 2 have two
qubits each and particle 3 has one qubit.}
\begin{ruledtabular}
\begin{tabular}{ccc}$n$ & $m$ & Distribution
\\ \hline\hline
2 & 2 & (1,1) \\
3 & 3 & (1,1,1) \\
4 & 2 & (2,2) \\
  & 4 & (1,1,1,1) \\
5 & 3 & (2,2,1) \\
  & 5 & (1,1,1,1,1) \\
6 & 2 & (3,3) \\
  & 3 & (2,2,2) \\
  & 4 & (2,2,1,1) \\
  & 6 & (1,1,1,1,1,1) \\
7 & 3 & (3,3,1), (3,2,2) \\
  & 4 & (2,2,2,1) \\
  & 5 & (2,2,1,1,1) \\
  & 7 & (1,1,1,1,1,1,1) \\
8 & 2 & (4,4) \\
  & 3 & (3,3,2) \\
  & 4 & (3,3,1,1), (3,2,2,1), (2,2,2,2) \\
  & 5 & (2,2,2,1,1) \\
  & 6 & (2,2,1,1,1,1) \\
  & 8 & (1,1,1,1,1,1,1,1) \\
\end{tabular}
\end{ruledtabular}
\end{table}


A corollary of the lemma is that there are no specific AVN
proofs in which one particle has more than $n/2$ qubits (this
result was used in Sec.~\ref{Sec:3}).


\subsection{AVN proofs with a minimum number of parties for any graph state}


Equipped with these tools, we can obtain all possible
distributions with a minimum number of particles allowing
specific AVN proofs for any graph state. We have obtained all
which are inequivalent under single-qubit unitary operations
for all graph states up to $n=8$ qubits. For this purpose, we
used the classification of graph states up to $n=7$ qubits
proposed in Ref.~\cite{HEB04} and the classification of
eight-qubit graph states proposed in Ref.~\cite{CLMP09}. Given
an $n$-qubit graph state, to obtain all the distributions
between a minimum number of parties allowing specific AVN
proofs we can use Table~\ref{Table1} in the following way.
Suppose that $n=6$. We first test whether AVN proofs are
possible for the simplest distributions permitted by
Table~\ref{Table1}, that is, $m=2$ parties with three qubits
each. If no AVN proof is possible, then we test whether there
are AVN proofs for the next possible distributions permitted by
Table~\ref{Table1}, that is, $m=3$ parties with two qubits
each, and so on.

Applying this method, we have obtained all inequivalent
distributions between a minimum number of particles for all
graph states with up to eight qubits. In the supplementary
material (see Footnote 2), we show all distributions between a
minimum number of particles for the $19$ classes of graph
states with up to six qubits, the $26$ classes of graph states
with seven qubits, and the $101$ classes of graph states with
eight qubits. In addition, we provide as supplementary material
(see Footnote) a computer program to obtain, given an $n$-qubit
graph state, all distributions between $m$ parties and all
distributions between a minimum number of parties which allow
AVN proofs.


\section{Conclusions}


We have developed tools with which to decide whether a
distribution of $n$ qubits between $m$ parties allows an
$m$-partite AVN proof specific for this distribution (i.e.,
which cannot be obtained using a state with fewer qubits). As a
result, we have obtained all inequivalent $m$-partite AVN
proofs using $n$-qubit $m$-particle quantum states with $n<9$
qubits and a minimum number $m$ of parties. This enables us to
obtain all inequivalent $m$-partite AVN proofs using $n$-qubit
$m$-particle quantum states with $n<9$ qubits with an arbitrary
number of parties.

The motivation of this work was to answer some natural
questions raised by recent experimental developments allowing
the preparation in the laboratory of graph states of several
particles, each carrying several qubits. The results presented
in this article provide tools to help experimentalists to
design tests of new AVN proofs and new Bell inequalities based
on these AVN proofs \cite{Mermin90a, Cabello05a, CGR08},
similar to those reported in Refs.~\cite{VPMDB07, CVDMC09} for
specific states but exploiting the possibility of
experimentally preparing new classes of graph states.


\section*{Acknowledgments}


The authors thank W.-B.~Gao, O.~G\"{u}hne, and A.~J.~L\'{o}pez
Tarrida for their useful comments and M.~Hein, J.~Eisert, and
H.~J.~Briegel for their permission for reproducing two figures
of Ref.~\cite{HEB04} in the supplementary material to this
article. The authors acknowledge support from Projects Nos.
P06-FQM-02243 and FIS2008-05596.


\newpage
\appendix

\section{Supplementary material}
\label{appendix}


\begin{table*}[h]
\caption{\label{Table2}Distributions of the qubits of a graph
state (numbered as in Fig.~\ref{Fig3}) between a minimum number
$m$ of particles $A,B,\ldots,F$ allowing AVN proofs specific
for the graph state. The table contains all inequivalent
distributions for all graph states up to six qubits.}
\begin{ruledtabular}
{\begin{tabular}{llllllll} ${\rm Graph\;state\;no.}$ & $m$
&$A$&$B$&$C$&$D$&$E$&$F$
\\ \hline \hline
$1$ & $2$&$ 1 $&$ 2 $& & & &\\
$2\;(GHZ_3) $&$ 3 $&$ 1 $&$ 2 $&$ 3 $& & & \\
$3\;(GHZ_4) $&$ 4 $&$ 1 $&$ 2 $&$ 3 $&$ 4 $& & \\
$4\;(LC_4)$&$ 2 $&$ 1,4 $&$ 2,3 $& & & & \\
$5\;(GHZ_5)$&$ 5 $&$ 1 $&$ 2 $&$ 3 $&$ 4 $&$ 5 $& \\
$6$&$ 3 $&$ 1 $&$ 2,4 $&$ 3,5 $& & & \\
$7\;(LC_5)$&$ 3 $&$ 1,5 $&$ 2,4 $&$ 3 $& & & \\
$8\;(RC_5)$&$ 3 $&$ 1 $&$ 2,5 $&$ 3,4 $& & & \\
$9\;(GHZ_6)$&$ 6 $&$ 1 $&$ 2 $&$ 3 $&$ 4 $&$ 5 $&$ 6 $ \\
$10$&$ 4 $&$ 1 $&$ 2,4 $&$ 5,6 $&$ 3 $& & \\
$11$&$ 3 $&$ 1,5 $&$ 2,3 $&$ 4,6 $& & & \\
$12$&$ 3 $&$ 1,5 $&$ 2,4 $&$ 3,6 $& & & \\
$13$&$ 2 $&$ 1,5,6 $&$ 2,3,4 $& & & & \\
$14\;(LC_6) $&$ 2 $&$ 1,4,5 $&$ 2,3,6 $& & & & \\
$15$&$ 3 $&$ 1,4 $&$ 2,6 $&$ 3,5 $& & & \\
$16$&$ 2 $&$ 2,3,4 $&$ 1,5,6 $& & & & \\
$17$&$ 2 $&$ 1,2,5 $&$ 3,4,6 $& & & & \\
$18\;(RC_6)$&$ 2 $&$ 1,2,4 $&$ 3,5,6 $& & & & \\
$19$&$ 2 $&$ 1,2,3 $&$ 4,5,6 $& & & & \\
\end{tabular}}
\end{ruledtabular}
\end{table*}


\begin{figure}[b]
\centerline{\includegraphics[width=0.9 \columnwidth]{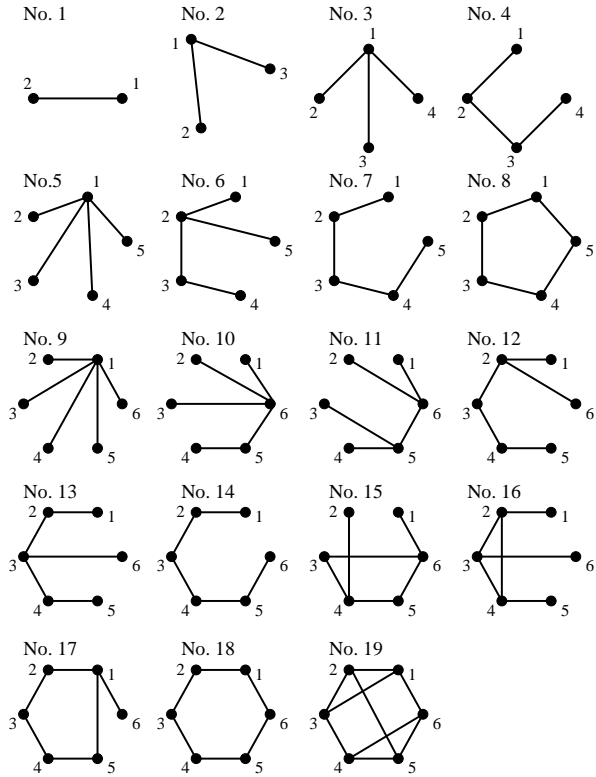}}
\caption{\label{Fig3}Graphs associated
to the $27$~classes on graph states with up to six qubits
inequivalent under local complementation and graph isomorphism.
Figure taken from Ref.~\cite{HEB04}.}
\end{figure}


\begin{table*}[htb]
\caption{\label{Table3}Distributions of the qubits of a graph
state (numbered as in Fig.~\ref{Fig4}) between a minimum number
$m$ of particles $A,B,\ldots,G$ allowing AVN proofs specific
for the graph state. The table contains all inequivalent
distributions for all seven-qubit graph states.}
\begin{ruledtabular}
{\begin{tabular}{lllllllll} ${\rm Graph\;state\;no.}$ & $m$
&$A$&$B$&$C$&$D$&$E$&$F$&$G$
\\ \hline \hline
$20\;(GHZ_7)$&$7$&$1$&$2$&$3$&$4$&$5$&$6$&$7$\\
21&$5$&$1,6$&$5,7$&$2$&$3$&$4$& & \\
22&$4$&$1,4$&$6,7$&$2,5$&$3$& & & \\
23&$4$&$1,4$&$6,7$&$2,5$&$3$& & & \\
24&$4$&$1,4$&$5,7$&$3,6$&$2$& & & \\
25&$3$&$1,4,6$&$2,5,7$&$3$& & & & \\
  &$3$&$1,4,6$&$5,7$&$2,3$& & & & \\
26&$3$&$1,4,6$&$5,7$&$2,3$& & & & \\
  &$3$&$1,3,5$&$4,6,7$&$2$& & & & \\
27&$3$&$1,4,5$&$2,3,6$&$7$& & & & \\
  &$3$&$1,4,5$&$2,3$&$6,7$& & & & \\
28&$3$&$1,3,6$&$2,5,7$&$4$& & & & \\
  &$3$&$1,3,7$&$2,4$&$5,6$& & & & \\
29&$3$&$1,3,4$&$2,6,5$&$7$& & & & \\
  &$3$&$1,3,4$&$2,6$&$5,7$& & & & \\
$30\;(LC_7)$&$3$&$1,5,7$&$2,6$&$3,4$& & & & \\
  &$3$&$1,3,5$&$2,4,6$&$7$& & & & \\
31&$4$&$1,4$&$2,6$&$3,5$&$7$& & & \\
32&$3$&$1,6,5$&$2,3,4$&$7$& & & & \\
  &$3$&$1,5,6$&$2,3$&$4,7$& & & & \\
33&$3$&$1,4,7$&$2,5,6$&$3$& & & & \\
  &$3$&$1,5,6$&$3,4$&$2,7$& & & & \\
34&$3$&$1,2,5$&$3,4,6$&$7$& & & & \\
  &$3$&$1,2,6$&$3,4$&$5,7$& & & & \\
35&$3$&$1,3,4$&$5,6,7$&$2$& & & & \\
  &$3$&$1,5,7$&$2,6$&$3,4$& & & & \\
36&$3$&$1,6,7$&$3,4,5$&$2$& & & & \\
  &$3$&$1,6,7$&$2,5$&$3,4$& & & & \\
37&$3$&$1,4,5$&$3,6,7$&$2$& & & & \\
  &$3$&$1,3,6$&$2,7$&$4,5$& & & & \\
38&$3$&$1,2,4$&$3,5,6$&$7$& & & & \\
  &$3$&$1,2,4$&$3,6$&$5,7$& & & & \\
39&$3$&$1,4,5$&$2,3,6$&$7$& & & & \\
  &$3$&$1,4,5$&$2,7$&$3,6$& & & & \\
$40(RC_7)$&$3$&$1,4,7$&$2,3,6$&$5$& & & & \\
  &$3$&$1,4,7$&$2,6$&$3,5$& & & & \\
41&$3$&$1,4,6$&$2,3,5$&$7$& & & & \\
  &$3$&$1,4,6$&$2,7$&$3,5$& & & & \\
42&$3$&$1,5,7$&$3,4,6$&$2$& & & & \\
  &$3$&$1,5,7$&$3,6$&$2,4$& & & & \\
43&$3$&$1,2,7$&$3,4,5$&$6$& & & & \\
  &$3$&$1,2,7$&$3,4$&$5,6$& & & & \\
44&$3$&$1,3,7$&$4,5,6$&$2$& & & & \\
  &$3$&$1,3,7$&$5,6$&$2,4$& & & & \\
45&$3$&$1,6,7$&$2,3,4$&$5$& & & & \\
  &$3$&$1,6,7$&$2,5$&$3,4$& & & & \\
\end{tabular}}
\end{ruledtabular}
\end{table*}


\begin{figure*}[htb]
\centerline{\includegraphics[width=0.9 \columnwidth]{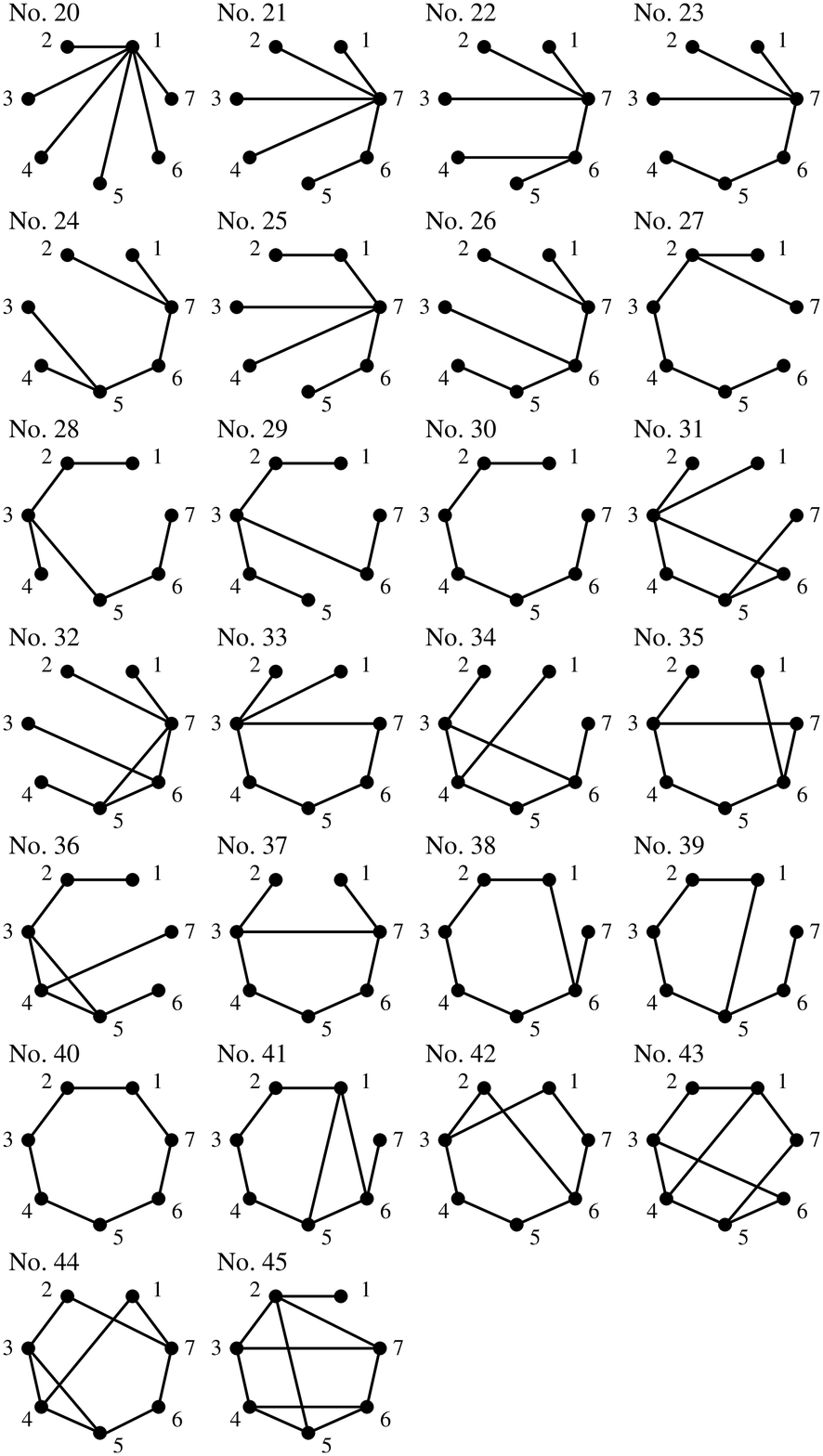}}
\caption{\label{Fig4}Graphs associated to the $22$~classes on
seven-qubit graph states inequivalent under local
complementation and graph isomorphism. Figure taken from
Ref.~\cite{HEB04}.}
\end{figure*}


\begin{table*}[htb]
\caption{\label{Table4}Distributions of the qubits of a graph
state (numbered as in Fig.~\ref{Fig4}) between a minimum number
$m$ of particles $A,B,\ldots,H$ allowing AVN proofs specific
for the graph state. The table contains all inequivalent
distributions for eight-qubit graph states nos. 46--90.}
\begin{ruledtabular}
{\begin{tabular}{llllllllll} ${\rm Graph\;state\;no}$ & $m$
&$A$&$B$&$C$&$D$&$E$&$F$&$G$&$H$
\\ \hline \hline
$46\;(GHZ_8)$& $8$ & $1$ & $2$ & $3$ & $4$ & $5$ & $6$ & $7$ &$8$ \\
47& $6$ & $1,6$& $2,7$ & $3$ & $4$ & $5$ & $8$ & & \\
48& $5$ & $1,5$ & $2,6$ & $3,7$ & $4$ & $8$ & & & \\
49& $5$ & $1,7$ & $3,5$ & $4,6$ & $2$ & $8$ & & & \\
50& $4$ & $ 1,6$ & $2,7$ & $3,5$ & $4,8$ & & & & \\
51& $4$ & $1,7$ & $2,5$ & $3,4$ & $6,8$ & & & & \\
52& $4$ & $1,5$ & $2,4$ & $3,6$ & $7,8$ & & & & \\
  & $4$ & $1,5,6$ & $3,4,7$ & $2$ & $8$ & & & & \\
  & $4$ & $1,5,7$ & $3,6$ & $4,8$ & $2$ & & & &\\
53& $4$ & $1,5$ & $2,6$ & $3,4$ & $7,8$ & & & & \\
  & $4$ & $1,4,6$ & $2,5$ & $3,7$ & $8$ & & & & \\
  & $4$ & $1,4,6$ & $5,7,8$ & $2$ & $3$ & & & & \\
54& $4$ & $1,7$ &$2,6$ & $3,5$ & $4,8$ & & & &\\
  & $4$ & $1,4,7$ & $2,6$ & $3,5$ & $8$ & & & &\\
  & $4$ & $1,5,6$ & $4,7,8$ & $2$ & $3$ & & & & \\
55& $3$ & $1,4,6$ & $2,3,5$ & $7,8$ & & & & & \\
56& $3$ & $1,4,6$ & $2,3,7$ & $5,8$ & & & & &\\
57& $3$ & $1,3,4$ & $2,5,7$ & $6,8$ & & & & &\\
58& $3$ & $1,3,6$ & $2,4,7$ & $5,8$ & & & & &\\
59& $3$ & $1,3,6$ & $4,7,8$ &$2,5$ & & & & &\\
60& $3$ & $1,5,7$ & $2,3,6$ & $4,8$ & & & & &\\
61& $3$ & $1,6,4$ & $2,3,5$ & $7,8$ & & & & &\\
62& $3$ & $1,5,7$ & $2,3,6$ & $8,4$ & & & & &\\
63& $3$ & $1,2,5$ & $3,6,8$ & $4,7$ & & & & &\\
64& $2$ & $1,2,4,7$ & $3,5,6,8$ & & & & & &\\
65& $2$ & $1,2,4,7$ & $3,5,6,8$ & & & & & &\\
66& $2$ & $1,2,4,7$ & $3,5,6,8$ & & & & & &\\
67& $2$ & $1,3,5,7$ & $2,4,6,8$ & & & & & &\\
$68\;(LC_8)$& $2$ & $1,4,5,8$ & $2,3,6,7$ & & & & & &\\
69& $4$ & $1,5$ & $2,6$ & $3,7$ & $4,8$ & & & &\\
70& $4$ & $1,4$ & $2,7$ & $3,6$ & $5,8$ & & & &\\
71& $4$ & $1,5,6$ & $2,4$ & $8,7$ & $3$ & & & &\\
  & $4$ & $1,5,6$ & $4,7,8$ & $3$ & $2$ & & & &\\
  & $4$ & $1,5$ & $2,4$ & $3,6$ & $7,8$ & & & &\\
72& $4$ & $1,5,7$ & $2,6$ & $4,8$ &$3$ & & & &\\
  & $4$ & $1,4,6$ & $5,7,8$ & $2$ & $3$ & & & &\\
  & $4$ & $1,5$ & $2,6$ & $3,7$ & $4,8$ & & & &\\
73& $3$ & $1,4,5$ & $2,3,6$ & $7,8$ & & & & &\\
74& $3$ & $1,5,7$ &$3,6,8$ & $2,4$ & & & & &\\
75& $3$ & $1,5,7$ & $2,4,6$ & $3,8$ & & & & &\\
76& $3$ & $1,2,4$ & $3,6,8$ & $5,7$ & & & & &\\
77& $3$ & $1,4,6$ & $2,3,5$ & $7,8$ & & & & &\\
78& $3$ & $1,6,7$ & $2,5,4$ & $3,8$ & & & & &\\
79& $3$ & $1,3,6$ & $2,4,7$ & $5,8$ & & & & &\\
80& $3$ & $1,3,6$ & $2,5,7$ & $4,8$ & & & & &\\
81& $3$ & $1,4,5$ & $2,3,6$ & $7,8$ & & & & &\\
82& $3$ & $1,3,7$ & $2,5,8$ & $4,6$ & & & & &\\
83& $3$ & $1,4,6$ & $2,5,7$ & $3,8$ & & & & &\\
84& $3$ & $1,3,5$ & $4,6,8$ & $2,7$ & & & & &\\
85& $3$ & $1,5,7$ & $3,4,8$ & $2,6$ & & & & &\\
86& $2$ & $1,3,6,7$ & $2,4,5,8$ & & & & & &\\
87& $2$ & $1,4,6,7$ & $2,3,5,8$ & & & & & &\\
88& $2$ & $1,2,4,7$ & $3,5,6,8$ & & & & & &\\
89& $2$ & $1,3,6,8$ & $2,4,5,7$ & & & & & &\\
90& $2$ & $1,4,6,8$ & $ 2,3,5,7$ & & & & & &\\
\end{tabular}}
\end{ruledtabular}
\end{table*}


\begin{table*}[htb]
\caption{\label{Table5}Distributions of the qubits of a graph
state (numbered as in Fig.~\ref{Fig4}) between a minimum number
$m$ of particles $A,B,\ldots,H$ allowing AVN proofs specific
for the graph state. The table contains all inequivalent
distributions for eight-qubit graph states nos. 91--146.}
\begin{ruledtabular}
{\begin{tabular}{llllllllll} ${\rm Graph\;state\;no}$ & $m$
&$A$&$B$&$C$&$D$&$E$&$F$&$G$&$H$
\\ \hline \hline
91& $2$ & $1,3,6,7$ & $2,4,5,8$ & & & & & &\\
92& $2$ & $1,3,5,7$ & $2,4,6,8$ & & & & & & \\
93& $2$ & $1,2,4,6$ & $3,5,7,8$ & & & & & & \\
94& $2$ & $1,3,5,8$ & $2,4,6,7$ & & & & & & \\
95& $2$ & $1,3,4,7$ & $2,5,6,8$ & & & & & & \\
96& $2$ & $1,3,5,7$ & $2,4,6,8$ & & & & & & \\
97& $2$ & $1,3,5,7$ & $2,4,6,8$ & & & & & & \\
98& $2$ & $1,3,5,6$ & $2,4,7,8$ & & & & & & \\
99& $2$ & $1,4,5,8$ & $2,3,6,7$ & & & & & & \\
$100\;(RC_8)$& $2$ & $1,3,6,8$ & $2,4,5,7$ & & & & & & \\
101& $3$ & $1,6,8$ & $2,3,5$ & $4,7$ & & & & & \\
102& $3$ & $1,3,7$ & $2,4,5$ & $6,8$ & & & & & \\
103& $3$ & $1,7,8$ & $2,3,5$ & $4,6$ & & & & & \\
104& $3$ & $1,3,7$ & $5,6,8$ & $2,4$ & & & & & \\
105& $3$ & $1,3,6$ & $2,4,5$ & $7,8$ & & & & & \\
106& $2$ & $1,3,4,6$ & $2,5,7,8$ & & & & & & \\
107& $2$ & $1,4,6,7$ & $2,3,5,8$ & & & & & & \\
108& $2$ & $1,4,6,8$ & $2,3,5,7$ & & & & & & \\
109& $2$ & $1,2,3,6$ & $4,5,7,8$ & & & & & & \\
110& $2$ & $1,2,5,6$ & $3,4,7,8$ & & & & & & \\
111& $2$ & $1,2,5,7$ & $3,4,6,8$ & & & & & & \\
112& $2$ & $1,3,4,6$ & $2,5,7,8$ & & & & & & \\
113& $2$ & $1,4,5,6$ & $2,3,7,8$ & & & & & & \\
114& $2$ & $1,3,5,6$ & $2,4,7,8$ & & & & & & \\
115& $2$ & $1,3,4,6$ & $2,5,7,8$ & & & & & & \\
116& $2$ & $1,2,3,5$ & $4,6,7,8$ & & & & & & \\
117& $2$ & $1,4,5,7$ & $2,3,6,8$ & & & & & & \\
118& $2$ & $1,5,8,6$ & $2,3,4,7$ & & & & & & \\
119& $2$ & $1,3,6,8$ & $2,4,5,7$ & & & & & & \\
120& $2$ & $1,2,4,8$ & $3,5,6,7$ & & & & & & \\
121& $3$ & $1,4,5$ & $2,7,8$ & $3,6$ & & & & & \\
122& $3$ & $1,5,7$ & $2,7,8$ & $3,4$ & & & & & \\
123& $2$ & $1,3,5,6$ & $2,4,7,8$ & & & & & & \\
124& $2$ & $1,4,6,7$ & $2,3,5,8$ & & & & & & \\
125& $2$ & $1,4,5,7$ & $2,3,6,8$ & & & & & & \\
126& $2$ & $1,2,4,6$ & $3,5,7,8$ & & & & & & \\
127& $2$ & $1,3,4,6$ & $2,5,7,8$ & & & & & & \\
128& $2$ & $1,3,6,7$ & $2,4,5,8$ & & & & & & \\
129& $2$ & $1,3,5,6$ & $2,4,7,8$ & & & & & & \\
130& $2$ & $1,2,5,8$ & $3,4,6,8$ & & & & & & \\
131& $2$ & $1,3,6,7$ & $2,4,5,8$ & & & & & & \\
132& $2$ & $1,4,5,6$ & $2,3,7,8$ & & & & & & \\
133& $2$ & $1,6,7,8$ & $2,3,4,5$ & & & & & & \\
134& $3$ & $1,4,6$ & $3,7,8$ & $2,5$ & & & & & \\
135& $2$ & $1,3,4,7$ & $2,5,6,8$ & & & & & & \\
136& $2$ & $1,3,4,5$ & $2,6,7,8$ & & & & & & \\
137& $2$ & $1,2,3,4$ & $5,6,7,8$ & & & & & & \\
138& $2$ & $1,3,4,6$ & $2,5,7,8$ & & & & & & \\
139& $2$ & $1,6,7,8$ & $2,3,4,5$ & & & & & & \\
140& $2$ & $1,4,5,6$ & $2,3,7,8$ & & & & & & \\
141& $2$ & $1,6,7,8$ & $2,3,4,5$ & & & & & & \\
142& $2$ & $1,3,7,8$ & $2,4,5,6$ & & & & & & \\
143& $2$ & $1,3,4,7$ & $2,5,6,8$ & & & & & & \\
144& $2$ & $1,2,5,6$ & $3,5,7,8$ & & & & & & \\
145& $2$ & $1,4,5,8$ & $2,3,6,7$ & & & & & & \\
146& $2$ & $1,2,4,5$ & $3,6,7,8$ & & & & & & \\
\end{tabular}}
\end{ruledtabular}
\end{table*}


\begin{figure*}[htb]
\centerline{\includegraphics[width=1.60
\columnwidth]{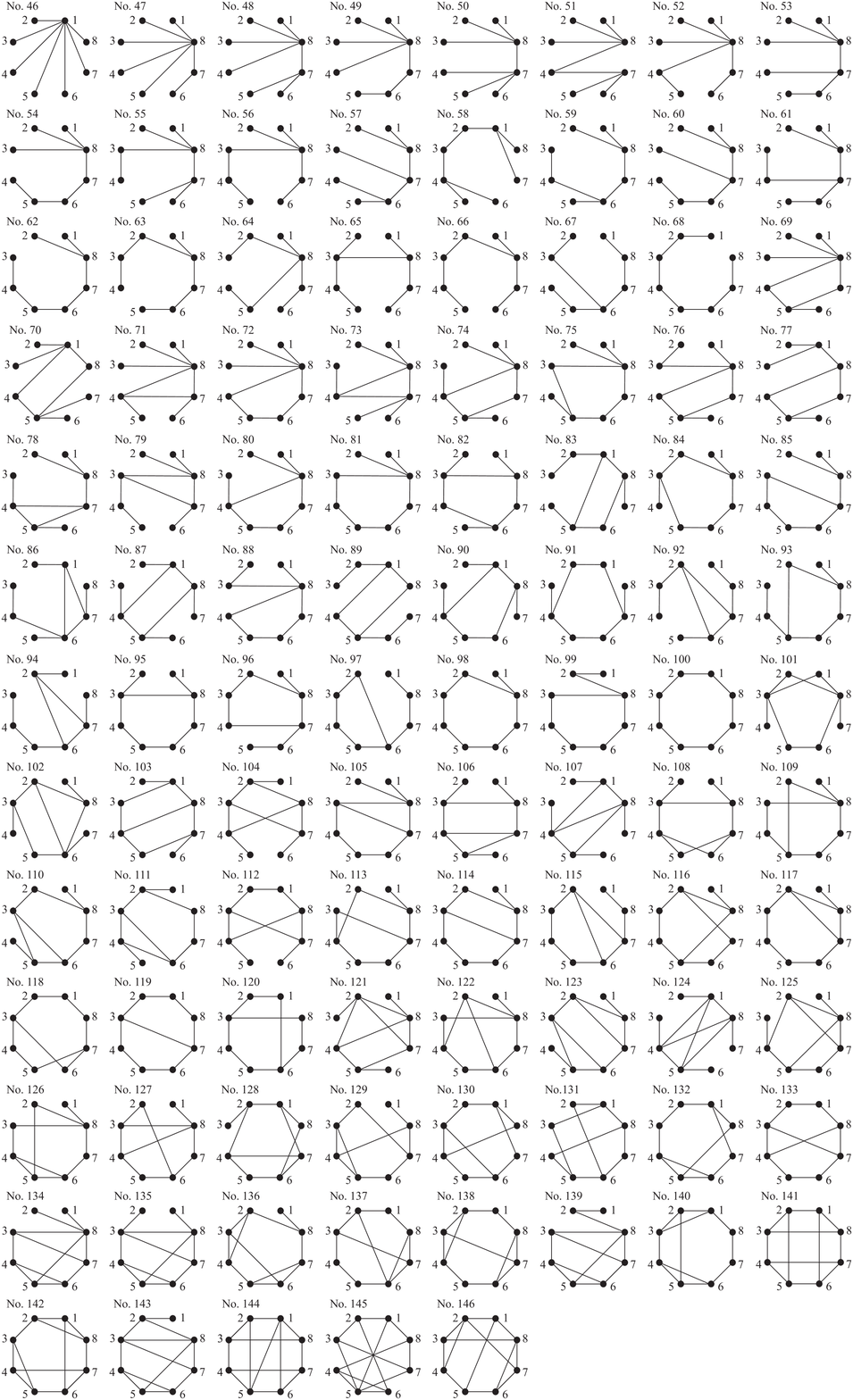}} \caption{\label{Fig5}Graphs
associated to the $101$~classes on eight-qubit graph states
inequivalent under local complementation and graph isomorphism.
Figure taken from Ref.~\cite{CLMP09}.}
\end{figure*}




\begin{thebibliography}{99}


\bibitem{EPR35}
A. Einstein, B. Podolsky, and N. Rosen,
Phys. Rev. {\bf 47}, 777 (1935).


\bibitem{Bell64}
J. S. Bell,
Physics (Long Island City, N.Y.) {\bf 1}, 195 (1964).


\bibitem{Mermin90b}
N. D. Mermin,
Phys. Rev. Lett. {\bf 65}, 1838 (1990).


\bibitem{HR83}
P. Heywood and M. L. G. Redhead,
Found. Phys. {\bf 13}, 481 (1983).


\bibitem{GHZ89}
D. M. Greenberger, M. A. Horne, and A. Zeilinger,
in {\em Bell's Theorem, Quantum Theory, and Conceptions of the
Universe}, edited by M. Kafatos (Kluwer Academic, Dordrecht,
1989), p. 69.

\bibitem{GHSZ90}
D. M. Greenberger, M. A. Horne, A. Shimony, and A. Zeilinger,
Am. J. Phys. {\bf 58}, 1131 (1990).

\bibitem{Mermin90a}
N. D. Mermin,
Phys. Rev. Lett. {\bf 65}, 3373 (1990).


\bibitem{Cabello01a}
A. Cabello,
Phys. Rev. Lett. {\bf 86}, 1911 (2001).

\bibitem{Cabello01b}
A. Cabello,
Phys. Rev. Lett. {\bf 87}, 010403 (2001).


\bibitem{Cabello05a}
A. Cabello,
Phys. Rev. Lett. {\bf 95}, 210401 (2005).

\bibitem{Cabello05b}
A. Cabello,
Phys. Rev. A {\bf 72}, 050101(R) (2005).


\bibitem{VPMDB07}
G. Vallone, E. Pomarico, P. Mataloni, F. De Martini, and V. Berardi,
Phys. Rev. Lett. {\bf 98}, 180502 (2007).


\bibitem{BLPK05}
J.~T.~Barreiro, N.~K.~Langford, N.~A.~Peters, and P.~G.~Kwiat,
Phys. Rev. Lett. {\bf 95}, 260501 (2005).

\bibitem{CVDMC09}
R. Ceccarelli, G. Vallone, F. De Martini, P. Mataloni, and A.
Cabello,
Phys. Rev. Lett. {\bf 103}, 160401 (2009).


\bibitem{GYXGCLPYCP09}
W.-B. Gao, X.-C. Yao, P. Xu, O. G{\"u}hne, A. Cabello, C.-Y. Lu,
C.-Z. Peng, T. Yang, Z.-B. Chen, and J.-W. Pan,
arXiv:0906.3390 (2009).


\bibitem{GLYXGGCPCP08}
W.-B. Gao, C.-Y. Lu, X.-C. Yao, P. Xu, O. G{\"u}hne, A. Goebel,
Y.-A. Chen, C.-Z. Peng, Z.-B. Chen, and J.-W. Pan,
Nat. Phys. doi:10.1038/nphys1603 (2010).


\bibitem{CM07}
A. Cabello and P. Moreno,
Phys. Rev. Lett. {\bf 99}, 220402 (2007).


\bibitem{SM}
 See supplementary material at
 http://link.aps.org/\linebreak
 supplemental/10.1103/PhysRevA.81.042110
 for a computer program in {\sc MATHEMATICA}.


\bibitem{Gottesman96}
D. Gottesman,
Phys. Rev.~A {\bf 54}, 1862 (1996).


\bibitem{VDM04}
M. Van den Nest, J. Dehaene, and B. De Moor,
Phys. Rev.~A {\bf 69}, 022316 (2004).

\bibitem{HEB04}
M. Hein, J. Eisert, and H. J. Briegel,
Phys. Rev.~A {\bf 69}, 062311 (2004).

\bibitem{CLMP09}
A. Cabello, A. J. L\'opez Tarrida, P. Moreno and J. R. Portillo,
Phys. Lett. A. {\bf 373}, 2219 (2009).


\bibitem{HDERVB06}
M. Hein, W. D\"{u}r, J. Eisert, R. Raussendorf, M. Van den Nest, and
H. J. Briegel,
 in {\em Quantum Computers, Algorithms and Chaos},
 edited by G. Casati, D.L. Shepelyansky, P. Zoller, and G. Benenti
 (IOS Press, Amsterdam, 2006).

\bibitem{DP97}
D. P. DiVincenzo and A. Peres,
Phys. Rev.~A {\bf 55}, 4089 (1997).

\bibitem{SASA05}
V. Scarani, A. Ac\'{\i}n, E. Schenck, and M. Aspelmeyer,
Phys. Rev.~A {\bf 71}, 042325 (2005).

\bibitem{CGR08}
A. Cabello, O. G\"{u}hne, and D. Rodr\'{\i}guez,
Phys. Rev.~A {\bf 77}, 062106 (2008).


\end{thebibliography}
\end{document}